\documentclass{jpp}
\usepackage{graphicx}
\usepackage{epstopdf, epsfig}
\usepackage{soul}

\shorttitle{Conditions for the onset of the current filamentation instability in the laboratory}
\shortauthor{N. Shukla et. al}

\title{Conditions for the onset of the current filamentation instability in the laboratory}

\author{N. Shukla\aff{1}\corresp{\email{nshukla@ist.utl.pt}},
  J.Vieira\aff{1}\corresp{\email{jorge.vieira@ist.utl.pt}},
  P. Muggli\aff{2},
  G. Sarri\aff{3},
  R. Fonseca\aff{1,4}\\
 \and L. O. Silva\aff{1}}

\affiliation{\aff{1}GoLP/Instituto de Plasmas e Fus\~ao Nuclear, Instituto Superior T\'ecnico, Universidade de Lisboa, Lisbon, Portugal
\aff{2}Max Planck Institute for Physics, Munich, Germany
\aff{3}Centre for Plasma Physics, School of Mathematics and Physics, Queen's University of Belfast, Belfast BT7 1NN, United Kingdom
\aff{4}DCTI/ISCTE, Instituto Universitario de Lisboa, Lisbon, Portugal}

\begin{document}

\maketitle

\begin{abstract}
Current Filamentation Instability (CFI) is capable of generating strong magnetic fields relevant to explain radiation processes in astrophysical objects and lead to the onset of particle acceleration in collisionless shocks. Probing such extreme scenarios in the laboratory is still an open challenge. In this work, we investigate the possibility of using neutral $e^{-}$ $e^{+}$ beams to explore the CFI with realistic parameters, by performing 2D particle-in-cell simulations. We show that CFI can occur unless the rate at which the beam expands due to finite beam emittance is larger than the CFI growth rate and as long as the role of competing electrostatic two-stream instability (TSI) is negligible. We also show that the longitudinal energy spread, typical of plasma based accelerated electron-positron fireball beams, plays a minor role in the growth of CFI in these scenarios. 
\end{abstract}

\section{Introduction}\label{intro}
The fireball is a promising model for the generation of $\gamma$-ray bursts (GRBs)~\citep{Cavallo-Rees-1978, Goodman-1986, Paczynski-1986, Rees-1992, Piran-1996, Piran-RMP-2004}. The model relies on the dissipation of kinetic energy of an ultrarelativistic flow, which emits $\gamma$-rays via synchrotron or synchrotron self-Compton emission. As a result, dense radiation and $e^{-}$ $e^{+}$ pair fluids are produced, known as a fireball \citep{Bahcall}. The interaction of the fireball beam, characterised by relativistic factors ranging from $10^2-10^6$, with the external medium can drive field structures that accelerate particles to high energies. As particles accelerate, they will also emit strong radiation bursts, with wavelengths ranging from $\gamma$-rays to radio waves. Astrophysical observations indicate that the main process leading to radiation emission is synchrotron radiation, which requires large amplitude magnetic fields on the order of Gauss to operate \citep{Piran-2005, Uzdensky-2014}. The origin of magnetic fields, and their amplification to these extreme values is a pressing challenge in astrophysics \citep{Widrow, Kronberg}. 

There has been an extensive effort, based on theoretical and numerical advances, with the objective of understanding the mechanisms by which strong magnetic fields are formed in astrophysical scenarios~\citep{Weibel-1959, Tzoufras-2006, Hantao-2015}. Medvedev, Waxman and Loeb \citep{Medvedev-ApJ-1999, Waxman-2009}  proposed that the current filamentation/Weibel Instability (CFI/WI) is a leading mechanism allowing for the growth of magnetic fields in the astrophysical context. The corresponding growth rates range from a few microseconds to a few tenths of a second, consistent with the time scales of GRBs~\citep{Schlickeiser-2005}. These instabilities arise due to  an anisotropic velocity distribution in the plasma (WI) or due to a counter-streaming flow of plasma slabs (CFI). Numerical calculations have shown that these instabilities are capable of generating strong magnetic fields with 10$^{-5}$ - 10$^{-1}$ of the energy density equipartition \citep{Luis-APJ-2003,Medvedev-APJ-2005, Dieckman, Stockem-2015,Fiuza-2012,Shukla-JPP-2010,Sayan-2011}.

Similar observational evidence of electron-positron ($e^{-}$ $e^{+}$) pair production has been found in TeV Blazers \citep{Gould-1967, Salamon-1998, Neronov-2009, PChang2012, Schlickeiser-2012}, where the interaction of TeV photons with the extragalactic background light produces ultra-relativistic $e^{-}$ $e^{+}$ beams. As these $e^{-}$ $e^{+}$ beams stream through the intergalactic medium (IGM), the collective beam-plasma instabilities can be relevant. The impact of beam-plasma instabilities upon $\gamma$-rays emission of bright TeV sources and their subsequent cosmological consequences have been previously investigated  theoretically and numerically using realistic parameters \citep{Schlickeiser-2013, Sironi-2014, Broderick-2014, PChang-2014, PChang-2016, Shalaby-2017}. The goal of this work is to identify conditions to explore such instabilities in laboratory conditions by using realistic finite size fireball beams. Leveraging on fully kinetic one-to-one particle-in-cell (PIC) simulations, we define the criteria for probing the Oblique Instability (OBI) and the CFI experimentally.

Exploring laboratory surrogates capable of reproducing these mechanisms under controlled conditions is a promising path to gain physical insights that would be otherwise inaccessible. One of the configurations that have been identified towards this goal is the study of the propagation of quasi-neutral relativistic fireball beams in the plasma \citep{Muggli-2013}. A globally neutral fireball beam is composed of equal amounts of electrons and positrons with identical density and spectral distributions. Recent experimental developments~\citep{Sari-Nature-2015} promise to make this exploration possible. The generation of quasi-neutral electron-positron fireball beams, with maximum energy $\simeq 400~\mathrm{MeV}$ (average $\mathrm{\gamma \sim 15}$), has been achieved in a laser-plasma accelerator. These beams have large energy spreads, they have a finite length and transverse size, and have limited charge. 

Another method for generating a fireball beam is to superimpose an electron $e^-$ and a positron $e^+$ bunches as could be done for example at SLAC \citep{Hogan-2010}. Numerical simulations show that this extremely relativistic fireball beam $\mathrm{\gamma = 40000}$ is also subject to CFI \citep{Muggli-2013}. Note that CFI of a mildly relativistic $e^-$ bunch $\mathrm{\gamma = 112}$ was observed showing filamentation and its coalescence \citep{Allen-2012}. Thus, although there have been efforts to understand the generation of magnetic fields through the Weibel/CFI under ideal conditions (i.e. infinitely wide planar plasma slabs) \citep{Frederiksen-ApJ-2004, Nishikawa-2009, Ricardo-POP-1999,Shukla-JPP-2012}, the role of realistic beam parameters in the growth of these instabilities remains to be understood in detail. 

In this work, we perform a detailed numerical and theoretical study of the interaction of a realistic fireball beam (with a length comparable or shorter than the plasma wavelength) in an uniform plasma using $\it{ab~initio}$ two-dimensional PIC simulations with the PIC code OSIRIS \citep{FonsecaR2002,Fonseca2008, Fonseca2013}. We examine in detail the temporal growth of the magnetic field that arises during the interaction between the fireball beam with the plasma. We then find that the growth of electrostatic modes, associated with competing instabilities, can be suppressed as long as the ratio between the beam density and the plasma density is sufficiently high. To make connection with recent experiments~\citep{Sari-Nature-2015}, we also investigate the role of the finite beam emittance in the beam dynamics, and find a threshold beam emittance for the occurrence of CFI. In addition, we found that the beam energy spread will not affect the growth of the CFI significantly. We consider ultra-relativistic fireball beams, with Lorentz factor $\gamma$ ranging between $\sim10^{3}-10^{4}$, propagating in the plasma with densities ranging between $10^{15}-10^{17}~\mathrm{cm}^{-3}$. These are parameters that can be explored in the laboratory. Our results show that the physics of OBI or CFI could be tested in the laboratory using presently or soon to be available electron-positron fireball beams.

\section{Simulations of the current filamentation instability}
\label{sec:illustrationofCFI}

The onset of CFI occurs when the ratio between the transverse beam size ($\sigma_y$) and the plasma skin depth ($k_p^{-1}=c/\omega_{p}$), is $k_p \sigma_y \geq 1$, where $\omega_{p} = \sqrt{4 \pi n_0 e^2 /m_e}$ is the plasma frequency, $n_0$ the background plasma density, $m_e$ the mass of the electron, $e$ the charge of the electron and c the speed of light. When the transverse beam size is larger than the plasma skin depth the plasma return currents can flow through the beam leading to the growth of CFI. If this condition does not hold, i.e when ($\sigma_y \leq c/\omega_{p}$), the CFI does not grow \citep{Roswell-1973, Su-1987, Chen-1987, Sentoku-2003, Blumenfeld-2007, Allen-2012}.

In order to illustrate the generation of magnetic fields through the CFI, we start by describing the results from 2D OSIRIS PIC simulations~\citep{FonsecaR2002, Fonseca2008, Fonseca2013}. The simulations use a moving window travelling at $c$. The simulation box has absorbing boundary conditions for the fields and for the particles in the transverse direction. The globally neutral fireball beam is initialized at the entrance of a stationary plasma with $n_0=10^{17}~\mathrm{cm}^{-3}$. The initial density profile for the electron and positron fireball beam is given by $n_b = n_{b0}\exp(-x^2/\sigma_{x}^2 - y^2/\sigma_y^2)$ where $n_{b0}=~n_0=10^{17}~\mathrm{cm}^{-3}$, $\sigma_{x}=0.99~c/\omega_p=10.2~\mu\mathrm{m}$ and $\sigma_y=2~c/\omega_p=20.4~\mu\mathrm{m}$ are the bunch peak density, length and transverse waist, respectively. The beam propagates along the $x$-axis with Lorentz factor $\gamma_b = 5.6 \times 10^4$, with transverse velocity spread $v_{th}/c = 1.7 \times 10^{-5}$ and with no momentum spread in the longitudinal direction. The simulation box dimensions are $L_{x}=~8.02~c/\omega_{p}$ and $L_{y}=~20.0~c/\omega_{p}$ with a moving window travelling at $c$ along x. The box is divided into $128 \times 512$ cells with $2\times2$ particles per cell.

\begin{figure}
\centerline{\includegraphics[width=1.0\textwidth]{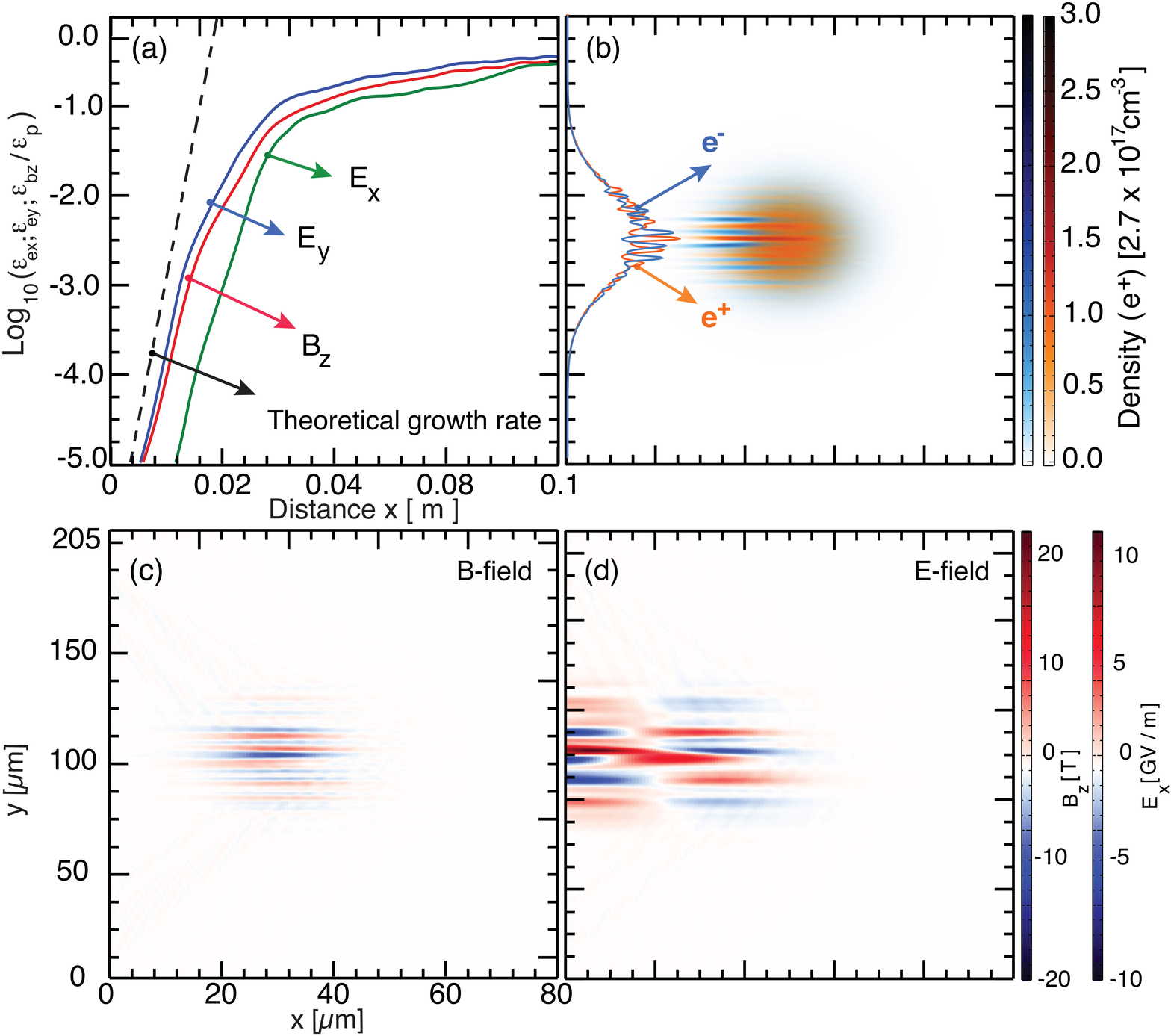}}
  	\caption{The interaction of a neutral e$^-$, e$^+$ fireball beam having a Gaussian profile with $\sigma_x = $  $2\sigma_y=20.4 \, \mu m $, peak density $n_b=2.7 \times 10^{15} \,\mathrm{m}^{-3}$, $\gamma_{b}=5.6 \times 10^4$ with a static plasma with $n_0 = n_b$.
  (\textit{a}) Evolution of Transverse magnetic $\epsilon_{bz}$ (red), Longitudinal $\epsilon_{ex}$ (green) and transverse electric $\epsilon_{ey}$ (blue) field energy as function of distance normalized to the initial kinetic energy of the beam $\epsilon_p = (\gamma_b-1)V$, where $V_b$ = $\pi \sigma_x \sigma_y$ is the volume of the beam. The dotted line represents the theoretical growth rate of the CFI. At time $t = \mathrm{1900.09}$ $\mathrm{[1/ \omega_p]}$, we show (\textit{b}) the density filaments corresponds to the electron e$^-$ (blue) and positron e$^+$ (red) spatially separated from each other (\textit{c}) the associated transverse magnetic ($B_z$) filaments at linear regime after 0.02 m (\textit{d})due to space charge radial electric field ($E_x$) created.}
\label{fig:1}
\end{figure}

Figure~\ref{fig:1} depicts the growth of the transverse magnetic field energy (panel a), the beam filaments due to the CFI (panel b), and the typical electromagnetic field structure (panels c and d). Figure~\ref{fig:1}a shows that the growth of the magnetic field energy as a function of the propagation distance is exponential as expected from the CFI. In Fig.~\ref{fig:1}a, the electromagnetic field is normalized with respect to the initial kinetic energy of the particles $\epsilon_p = (\gamma_b - 1)V_b$, where $V_b = (\pi\sigma_x\sigma_y) $ is the volume of the beam. Simulations reveal that the field energy grows at the expense of the total kinetic energy of the fireball beam. The linear growth rate of the CFI measured in the simulation is $\Gamma_{CFI}/\omega_{p} \simeq 6.0 \times 10^{-3}$, in good agreement with Silva et al 2002 \citep{Luis-2002}. As a consequence of the instability, the beam breaks up into narrow (with a width on the order of $0.5\,\mathrm{c/\omega_p}$, which correspond to 5$\, \mathrm{\mu m}$ for our baseline paramters) and high current density filaments. Figure~\ref{fig:1}(b) shows that these electron-positron filaments are spatially separated from each other. Each filament carries strong currents which lead to the generation of strong out-of-the-plane (i.e. azimuthal) magnetic fields with amplitudes beyond 20 T. The azimuthal magnetic fields are also filamented, as shown in Figure~\ref{fig:1} (c). Because of their finite transverse momentum, simulations show that current filaments can merge. As merging occurs, the width of the filaments increases, until beam breakup occurs. At this point, the CFI stops growing, and no more beam energy flows into the generation of azimuthal magnetic fields. Simultaneously, radial E-fields above 10 GV/m are also generated~(Fig.~\ref{fig:1}d).

\section{Role of the peak beam density and beam duration in the growth of current filamentation instability}
\label{sec:longerbeam}
In the previous section, and for illustratation purposes only, we have considered that the total beam density was twice the background plasma density. In this section, we will investigate the propagation of beams with lower peak densities. In order to keep the number of particles constant, we then increase the beam length, such that $\sigma_x \geq \lambda_{p}$. In these conditions, the OBI competes with the CFI \citep{Bret-POP-2009}. The OBI can grow  when the wave-vector is at an angle with respect to the flow velocity direction, and it leads to the generation of both electric and magnetic field components. The maximum growth rate for the CFI and the OBI are given by~$\Gamma_{CFI} \sim \sqrt{\alpha/\gamma_b}\, \beta_{b0}$ and $\Gamma_{OBI} \sim \sqrt{3}/2^{4/3} (\alpha/\gamma_b)^{1/3}$ respectively ~\citep{Bret-2006}, where $\alpha$ is the beam ($n_b$) to plasma density ($n_0$) ratio and $\beta_{b0} = v_b/c$ is the normalized velocity of the beam. Thus, the ratio between the CFI growth rate and the OBI growth rate, which is given by:
\begin{equation}
\label{eq:ratio-obi-cfi}
	\frac{\Gamma_{OBI}}{\Gamma_{CFI}} = \frac{\sqrt{3}}{2^{4/3}} \frac{1}{\beta_b} \left( \frac{\gamma_b}{\alpha} \right)^{1/6}
\end{equation}

Equation (3.1) provides the range of parameters for which each instability will dominate. The OBI dominates over the CFI when $\alpha$ is smaller 1. For the number of particles we considered in the simulations, this implies that only beams with $\sigma_y < 2 \lambda_{pe}$ will be subject to the CFI.

In order to verify this hypothesis, we have carried out additional two-dimensional OSIRIS PIC simulations using the initial set up described in Sec.~\ref{sec:illustrationofCFI}, varying $\sigma_x$ between $2\lambda_p$ and $10\lambda_p$, for which $\alpha$ varies between $0.0026$ and $1.0$. In all these cases, our results have consistently shown the evidence of the OBI growth.

\begin{figure}
  \centering
    \includegraphics[width=1.0\textwidth]{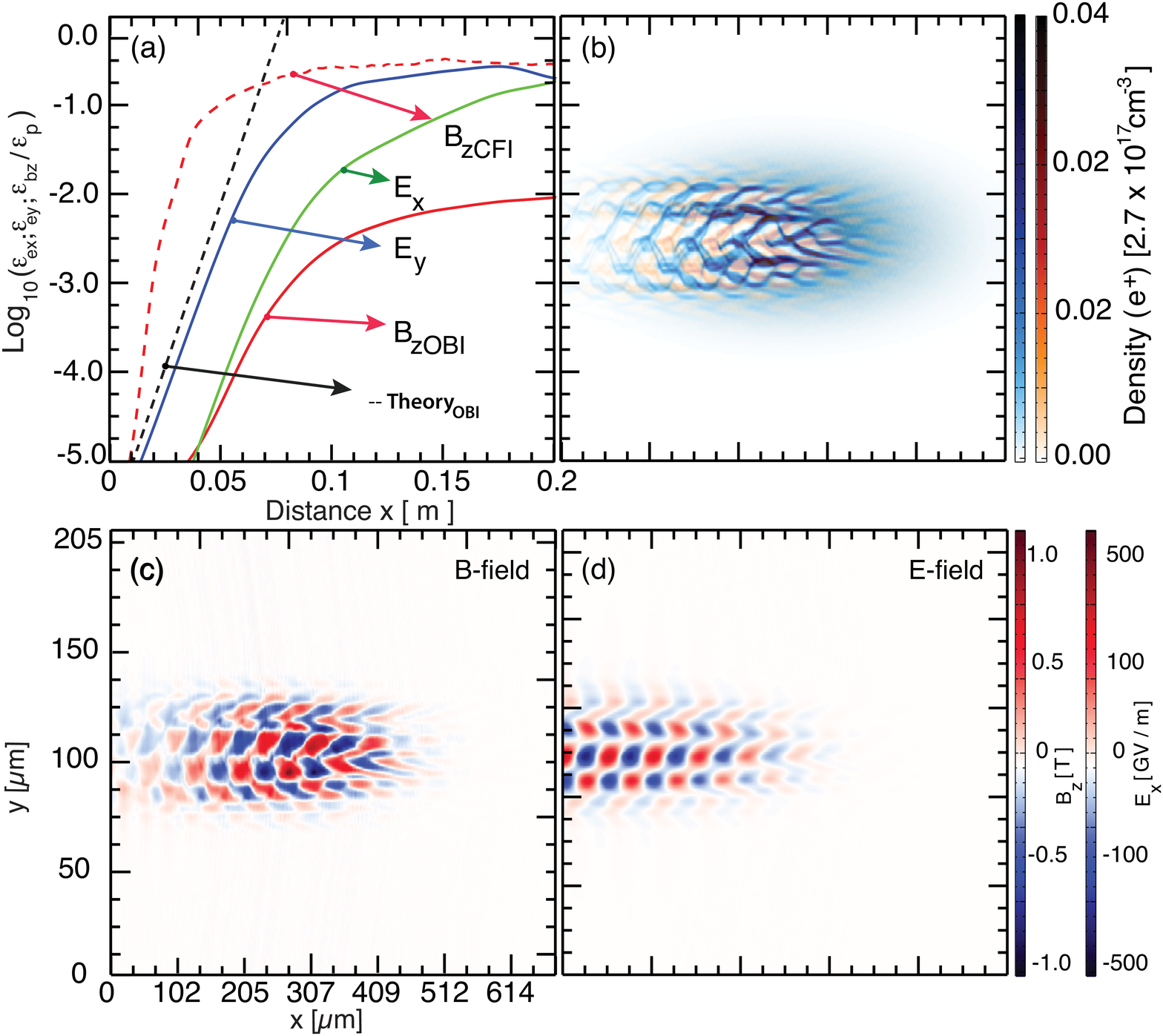}
  \caption{Interaction of a neutral e$^-$, e$^+$ fireball beam with longitudinal size  $\sigma_x = 2 \lambda_{p}$ and a static plasma by keeping constant beam particle number. (\textit{a}) Evolution of transverse magnetic $\epsilon_{bz}$ (red), longitudinal $\epsilon_{ex}$ (green) and transverse electric $\epsilon_{ey}$ (blue) field energy as function of distance normalized to the initial kinetic energy of the beam $\epsilon_p$. The dotted line represents the theoretical growth rate of OBI. Dashed red-line is the evolution of the out of the plane magnetic field for the condition of the simulation shown in Fig. 1(a). At time $t = \mathrm{8880.07}$ $\mathrm{[1/ \omega_p]}$, we show (\textit{b}) the density filaments corresponds to the electron e$^-$ (blue) and positron e$^+$ (red) spatially separated from each other (\textit{c}) the associate transverse magnetic ($B_z$) filaments at linear regime between $x_1 = 0.0551- 0.0556$ $\mathrm{m}$ (\textit{d}) the space charge separation leads to radial electric field ($E_x$).}
  \label{fig:2}
\end{figure}

In Figure~\ref{fig:2}, we show an illustrative simulation result considering $\sigma_x = 2\lambda_{p}$, with $n_b = 1.274\times10^{15}~ \mathrm{cm^{-3}}$, for which $\alpha= 0.01274$. In order to describe the propagation of a longer beam, we have increased the simulation box length. We then increased the longitudinal box length to $L_x = \mathrm{63}\,c/\omega_{p}$ ($L_y = \mathrm{20}\,c/\omega_{p}$ remains identical to that of Sec.~\ref{sec:illustrationofCFI}). The box is now divided into $\mathrm{1024} \times \mathrm{512}$ cells with $2\times2$ particles per cell for each species.

Figure~\ref{fig:2} (a) illustrates the evolution of the longitudinal and transverse electric and transverse magnetic energy (normalized to $\epsilon_p = (\gamma_b - 1)V_b$, where $V_b = (\pi \sigma_x \sigma_y) $ is the volume of the beam). The emergence of oblique modes can be seen in Fig.~\ref{fig:2} (b), which shows tilted beam filaments. In a multi-dimensional configuration, the oblique wave-vector couples the transverse (filamentation) and longitudinal (two-stream) instabilities resulting in the electromangetic beam plasma instability. Unlike Fig.~\ref{fig:1}, the simulation results in Fig.~\ref{fig:2} show that the transverse electric field ($E_y$) component provides the dominant contribution to the total field energy. The plasma is only weakly magnetized $\omega_c/\omega_p = 0.01$, much lower than in Fig.~\ref{fig:1}c where $\omega_c/\omega_p\simeq 0.6$, and the CFI does not play a critical role in the beam propagation. The longitudinal and transverse electric fields grow exponentially, as predicted by the linear analysis of the OBI, matching well the simulation results. The growth rate measured in the simulations is $\Gamma_\mathrm{{max}}/\omega_{p} \simeq \Gamma_\mathrm{{OBI}} \simeq 2.1 \times 10^{-3}$, while the theoretical growth rate is $\simeq 2.0 \times 10^{-3}$. The OBI generates plasma waves with strong radial electric fields in excess of 500 MV/m [Fig.~\ref{fig:2} (d)]. After 20 cm, the OBI saturates. Despite being limits of the same instability, the electromagnetic beam plasma instability, we will refer to the CFI and the OBI has manifestition of qualitative different behaviour of the same instability.  

\section{Effects of finite beam waist and emittance}
\label{sec:emittance}
Theoretical and numerical studies performed to identify the effect of beam emittance on the growth of plasma instabilities and their saturation \citep{Luis-2002, Ricardo-POP-1999, Shukla-JPP-2012} typically assume that the beam is infinitely wide. In this section, we will investigate the role of the beam emittance considering finite beam size effects, in order to make closer contact with laboratory conditions. To study the influence of the beam emittance on the propagation, we first consider the equation for the evolution of the beam waist $\sigma_y$ in vacuum \citep{Schroeder-2011}
\begin{equation}
\label{eq:include_label_here}
\frac{1}{c^2}\frac{\mathrm{d}^2 \sigma_y}{\mathrm{dt^2}} = \frac{\epsilon_N^2}{\sigma_y^3 \gamma_b^2},
\end{equation}
where $\sigma_y$ is the beam radius, $\epsilon_N \simeq \mathrm{\Delta p_{\perp}}\sigma_y$ is a figure for the beam emittance (corresponding to the area of the beam transverse phase space), and $\mathrm{\Delta p_{\perp}}$ is the transverse momentum. According to Eq. (\ref{eq:include_label_here}), the evolution for $\sigma_y$ and for sufficiently early times is given by:
\begin{equation}
\sigma_y \simeq \sigma_{y0} \left(1 + \frac{\epsilon_N^2~t^2 c^2}{2 \sigma_{y0}^4 \gamma_b^2} \right)^{1/2}~,
\end{equation} 
where $\sigma_{y0}$ is the initial beam radius. Hence, according to Eq.(4.1), the rate at which $\sigma_y$ increases is:
\begin{equation}
\frac{1}{\sigma_y} \frac{\mathrm{d}\sigma_y}{\mathrm{dt}} = \frac{t~c^2~\epsilon_N^2}{\sigma_{y0}^2 \gamma_b^2}~,
\end{equation}
Equation (4.3) indicates that the beam expands in vacuum due to its transverse momentum spread. As the beam expands, $n_b$ decreases as $n_b/n_0 \sim (\sigma_{y0}/\sigma_y)^2$, in 3D, and as $(\sigma_{y0}/\sigma_y)$, in 2D. Because of the reduction of $n_b/n_0$, the growth rates for the CFI and for the OBI will also decrease. We then estimate that these instabilities (i.e CFI and OBI) are suppressed when the rate at which $n_b/n_0$ decreases is much higher than the instability growth rate. Matching the rate at which the beam density drops, which in 2D is given by $(1/\sigma_y)~(\mathrm{d} \sigma_y/\mathrm{d t})$, to the growth rate of the instability ($\Gamma$) gives an upper limit for the maximum beam divergence $\theta = \mathrm{\Delta p_{\perp}}/\gamma_b$ (and emittance $\epsilon_N \approx \sigma_r (<p_{\perp}^2>)^{1/2}$) allowed for the growth of the CFI/OBI:
\begin{equation}\label{eq1}
\theta = \left( \frac{\Gamma \sigma_{y0}^2}{L_{\mathrm{growth}}~c} \right)^{1/2},
\end{equation}
where we have considered that $t\sim L_{\mathrm{growth}}/c$ in Eq. (4.4), being $L_{\mathrm{growth}}$ the growth length of the CFI/OBI instability. Equation (4.4) then gives the threshold beam divergence, beyond which the CFI/OBI will be suppressed. It indicates that beams with higher energy can support higher divergences and still be subject to the growth of the CFI because the beam expands slowly in comparison to lower energy beams. Similarly, beams with higher $\sigma_{y0}$ also support higher emittance than narrower beams because of the slower expansion rate.

To confirm our theoretical findings, we performed additional two-dimensional simulations using fireball beams with relativistic factors $\gamma_b =700, 1050, 1400$ (the lower $\gamma_b$ factors used now, in comparison to Sec. 2, minimize the computational requirements). We use $\sigma_{x}=0.22~c/\omega_p=11.7~\mu\mathrm{m}$ and $\sigma_{y}=10~c/\omega_p=530~\mu\mathrm{m}$ with peak density $n_{b0}=10~n_0=10^{15}~\mathrm{cm}^{-3}$. For each case, we varied the transverse temperature $ \mathrm{\Delta p_{\perp}} = \gamma_b \theta_{ze0} = 1, 3, 5, 7, 10$ and $20$ in order to determine the threshold beam spread for the occurrence of instability. We note that we have used the classical addition of velocities in the beam thermal spread initialization in order to more clearly identify the dependence of evolution of the instabilities with emittance.

\begin{figure}
  \centering
    \includegraphics[width=1.0\textwidth]{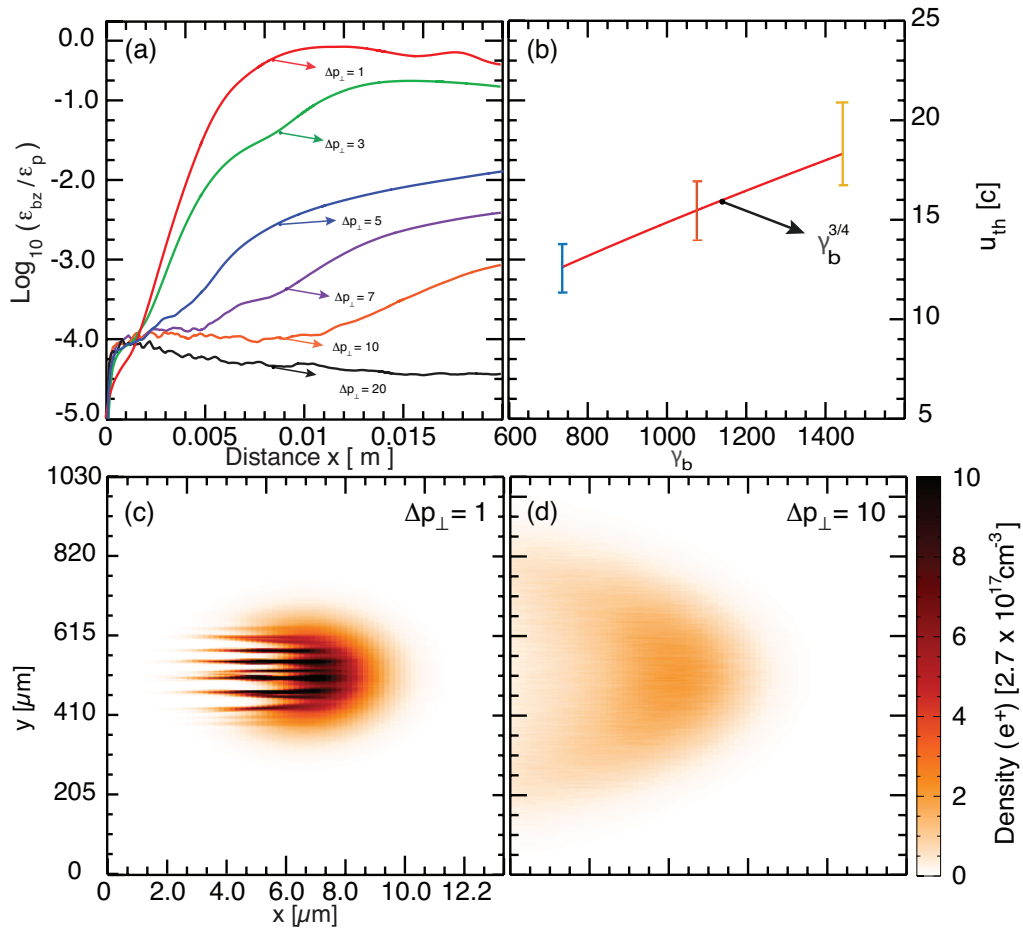}
  \caption{(\textit{a}) Temporal evolution of the transverse magnetic field energy for different beam emittance (\textit{b}) Thermal velocities as function of Lorentz factor $\gamma_b$, filamentation suppressed for higher thermal velocities. At time $t = \mathrm{705.60}$ $\mathrm{[1/\omega_p]}$, panels (\textit{c}) -(\textit{d})  show beam filaments for thermal velocities $ \mathrm{\Delta p_{\perp}} = 1,3,5,7,10,20$ from 2D PIC simulations.}
  \label{fig:3}
\end{figure}

Figure~\ref{fig:3}(a) shows that the magnetic field energy decreases with increasing transverse momentum spread. Fig. \ref{fig:3} (a) also shows a transition in the evolution of the magnetic energy between $ \mathrm{\Delta p_{\perp}} = 10$, where the B-field still grows at the end of the simulation, and $ \mathrm{\Delta p_{\perp}} = 20$, where the B-field decreases with propagation distance. According to Eq. (\ref{eq1}), using $L_{\mathrm{growth}} \sim 0.06$ $\mathrm{m} $ and $\Gamma_{\mathrm{CFI}} \sim 1.657 \times 10^{11} s^{-1}$, we obtain the threshold $\theta \sim 0.12$ for the shutdown of the instability. This is in good agreement with Fig. \ref{fig:3} (a). 

Figure~\ref{fig:3}~(b) depicts the dependence of the threshold beam emittance with the fireball beam energy. Figures~\ref{fig:3}~(c)-(d)  show the positron density for two simulations, where all the parameters are kept constant, except for the beam emittance. In particular, in Fig.~\ref{fig:3}~(c) a beam emittance of $\mathrm{\Delta p_{\perp}} = 1$, much smaller that the threshold value given by Eq. (3), has been considered. In this case the CFI develops, leading to the filamentation of the beam (see Fig.~\ref{fig:3}c) and to the exponential growth of the magnetic field energy (see Fig.~\ref{fig:3}~(a), red curve). However, in the second case (Fig.~\ref{fig:3}~(d)) a higher beam emittance $\mathrm{\Delta p_{\perp}} =10$  is considered. This suppresses the growth of the magnetic field energy (see Fig.~\ref{fig:3}~(a), black curve). As a result, the beam expands before the development of the CFI. These results show that the growth of CFI can only be achieved if the beam emittance is sufficiently small.

\section{Effect of beam energy spread}
\label{sec:energyspread}
In typical laboratory settings \citep{Sari-Nature-2015}, electron-positron fireball beams can contain finite energy spreads. It is, therefore, important to evaluate the potentially deleterious role of the energy spread in the growth of CFI. In this section, we then present simulation results with finite longitudinal momentum spreads. We consider that the central beam relativistic factor is $\gamma_b = 700$, and compare two simulations with $\Delta p_x/\gamma_b = 0.13$ and $\Delta p_x/\gamma_b = 0.29$ ($\Delta p_x$ is the longitudinal momentum spread). All other simulation parameters are similar to those described in Sec. 4.

\begin{figure}
  \centering
    \includegraphics[width=1.0\textwidth]{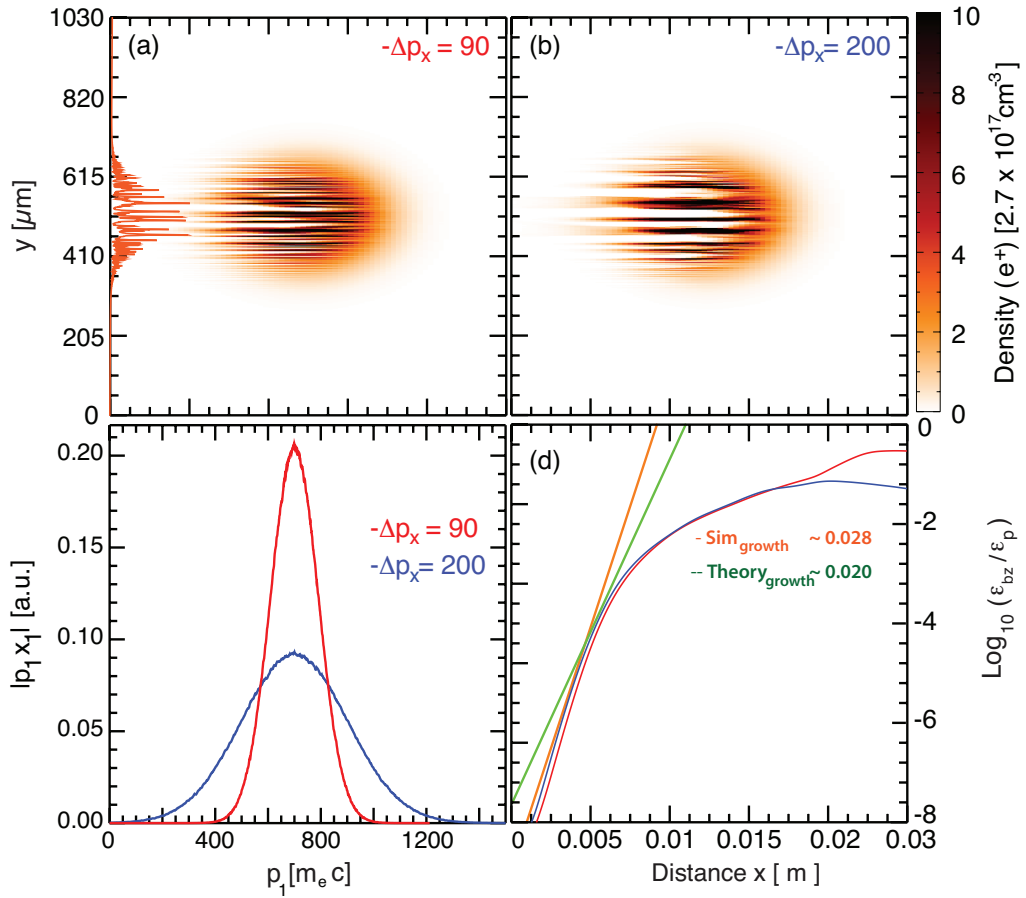}
  \caption{(\textit{a}) -(\textit{b}) Temporal evolution ($t = \mathrm{705.60}$ $\mathrm{[1/ \omega_p]}$ ) of positron density exhibits the filaments for energy spread $\mathrm{\Delta p_x} = \mathrm{90, 200}$ (\textit{c}) (Color) the spectrum of two different energy spread (\textit{d}) Growth rate of CFI for different energy of beam.}
  \label{fig:4}
\end{figure}

Figures (4) (a)-(b) show the temporal evolution of the beam electrons density for $\mathrm{\Delta p_x/\gamma_b} = 0.13$ (Fig. 4(a))~ and for $\mathrm{\Delta p_x/\gamma_b}= 0.29$ (Fig. 4(b)). The initial energy spectra of these two beams are shown in Fig.~4(c). Figure 4~(d) shows the comparison of the magnetic field energy evolution.  The blue curve shows the growth of magnetic field energy generated by the fireball beam with energy spread $\mathrm{\Delta p_x/\gamma_b} = 0.29$, while the red curve is associated with the lower energy spread $\mathrm{\Delta p_x/\gamma_b} = 0.13$. Simulation results demonstrate that the CFI grows in all cases, in agreement with analytical calculations~\citep{Inglebert-2012}. The green line in Fig.~4(d) is the theoretical growth rate $\Gamma/\omega_p \simeq 2.0\times 10^{-2}$, which is in good agreement with the simulation growth rate (shown by the red line in Fig.~4b).
\section{Summary and conclusions}
\label{sec:conclusions}
In summary, the growth and saturation of a ultra-relativistic beam propagating through a plasma have been investigated using particle-in-cell (PIC) simulations. We have shown that short fireball beams, i.e beams shorter than the plasma wavelength, interacting with uniform plasmas lead to the growth of the CFI. For typical parameters available for experiments, the instability can generate strong transverse magnetic field on the order of the MGauss. The instability saturates after 10 cm of propagation in a plasma with $n_0 \sim 10^{17}~\mathrm{cm}^{-3}$.

We have demonstrated that the beam density needs to be higher than the background plasma density to suppress the growth of the competing OBI instability, which leads to the growth of electrostatic modes (instead of electromagnetic). Beams with lower peak densities will then drive the OBI, which results in tilted filaments and the generation of mostly electrostatic plasma waves. We have also showed that the beam emittance needs to be minimized, reducing transverse beam defocusing effects, which can shutdown the CFI or the OBI if the beam defocuses before these instabilities grow. We have also extended our numerical studies to investigate the effect of finite fireball energy spreads on the growth of CFI, and showed that the energy spreads of currently available fireball beams allow for the growth of CFI in the laboratory.

In conclusion, we have identified the factors for the generation of strong magnetic fields via CFI. We expect that the results will influence our understanding of astrophysical scenarios, by revealing the laboratory conditions where these effects can be studied.

\acknowledgements
This work was partially supported by the European Research Council (ERC-2016-InPairs 695088). Simulations were performed at the IST cluster (Lisbon, Portugal). J. V. acknowledges the support for FCT (Portugal).

\bibliographystyle{jpp}

\bibliography{jpp-instructions}

\end{document}